\documentclass{article}
\usepackage{amsmath,amssymb}
\usepackage{graphicx}
\newcommand{\op}[1]{\hat{#1}}
\newcommand{\ket}[1]{| #1 \rangle}
\newcommand{\bra}[1]{\langle #1 |}

\title{Quantum ``Pathfinder''\\
\large\sf (Extended Abstract)}
\date{}
\author{Alexander Yu.\ Vlasov}
\begin{document}
\maketitle
\begin{abstract}
 It is discussed an opportunity to introduce new class of quantum algorithms 
based on possibility to express amplitude of transition between two states of 
quantum system as sum of some function along all possible classical paths. Continuous
analogue of the property with integral on all possible paths is well known due 
to Feynman approach and ensures the correspondence with classical minimal action 
principle. It is less technical and rather popular introduction to earlier work
about application of Lagrangian methods in quantum computing \cite{QLagr}.
\end{abstract}

\centerline{* * *}

\begin{flushright}
\footnotesize
{\em
{\rm Alice:} Our main task --- is practical realization of\\
 a general purpose quantum computer.\\
\smallskip
{\rm Bob:} I have got one, {\sffamily PQ1} 64Qb $\sqrt{2}$GHZ. And so?\\
\smallskip
{\rm Alice:} Excellent! Try Shor's factoring algorithm. \\
\smallskip
{\rm Bob:} I am trying: $55=7 \times 9\frac{1}{7} ?\ \ldots\ 8 \times 6\frac{7}{8}?$\\
}
\smallskip
(Quantum Communications, \today)
\end{flushright}

Lack of algorithms is certain problem of quantum information science.
Possibly, famous Shor's factoring algorithm would be enough, but even if we 
would have general purpose 
quantum computer with hundreds qubits, overcome decoherence, {\em etc.,} 
it still may fail with decomposition of number 55 simply because some quantum 
phase gate instead of $e^{32i\phi}$ is doing $e^{31.9 i\phi}$. 

Formally, it may be anyway general purpose quantum computer, because there is
implemented universal set of quantum gates; in principle the system may perform 
arbitrary unitary operator in exponentially big Hilbert space and be quite
appropriate for wide variety of tasks like simulation of quantum system
or realization currently unknown quantum algorithms.

The problem discussed above is simply usual declaration about difficulty
to realize some ``pure digital'' tasks on analogue element. The quantum 
computer is using methods both discrete and analogue processing and it 
is useful to find good balance.

Class of NP-complete tasks is even more powerful than Shor's factoring algorithm.
Well known example, {\em traveling salesman problem} (TSP), looks ``more physicaslly,'' 
because it is a task about finding shorter paths. Really one of earliest paper about 
quantum algorithm was devoted just the TSP \cite{Cer}.
It should be mentioned, that problem of finding {\em minimal} path in graph 
\cite{Cer} really even beyond scope of NP-complete problems \cite{LaurAI}, 
because formally it differs from initial TSP, {\em i.e.} about finding a 
path {\em shorter than given length}.

Unfortunately, the solution \cite{Cer} need for exponential number of particles,
but the problem of minimal path anyway devote attention, due to very
natural bond between physical and combinatorial tasks displayed in such
kind of problems. There is well known and quite 
popular explanation by Feynman \cite{QED}, why laws of quantum mechanics cause 
photon to travel between two points by shortest possible way\footnote{Even in 
nonhomogeneous media, including discontinuous case like refraction on boundary 
of air and water.} in agreement with classical minimal action principle. 

However, search for new quantum algorithms of such a kind have been
failing, C. Bennett wrote in \cite{Bennett} 
\begin{quote}
\small
``No fast quantum algorithms have been found for other famous search or
optimization problems, such as the traveling salesman problem and the
large class (called NP-complete) of problems equivalent to it. These
problems, like the naive approach to factoring, can be cast as searches
for a successful solution among exponentially many candidates; but
unlike factoring, no way is known of transition them into problems
with a periodic structure amenable to detection by quantum interference.''
\end{quote}

In the Feynman's approach the photon just travelling along all possible classical
paths, and due to oscillating term $\exp({\frac{i}{\hbar}\mathcal A})$ only 
trajectories near extremum of action function $\mathcal A$ survive. 
The Bennett's formulation above looks very close to the principle.
Why does not try to use the sums along all paths for some quantum
optimization algorithm?

Rigorously, we should not expect precisely solution of TSP-like problems
for any initial data using the method ---  there is {\em diffraction} phenomenon, 
{\em i.e.,} if irregularities of media are comparable with period of spatial 
oscillations (wave length), instead of one trajectory we may have some periodic
angular scaterring. But even such phenomenon may be useful, say Simon and Shor quantum 
algorithms in such picture may be compared with finding step of {\em diffraction 
grating} by spectral decomposition of falling white light (or X-ray analysis of 
crystals).

Anyway such phenomenon devotes applications or research in area of quantum 
information science. One obvious difficulty --- is usage finite dimensional 
spaces in theory of quantum computing. This difficulty was partially overcome
in \cite{QLagr}, there was shown, that most methods used in continuous theory may
be rewritten for finite-dimensional discrete case using quite straightforward ideas. 

The second problem is misconceptions due to formal difficulty with situating the 
task in classification of quantum algorithms. Say, it was proven in \cite{BBBV97},
that 
\begin{quote} 
\small
``\ldots  that relative to an oracle chosen uniformly at random, with 
probability 1, the class {\bf NP} cannot be solved on a quantum Turing machine 
in time $o(2^{n/2})$. We also show that relative to a permutation oracle chosen
uniformly at random, with probability 1, the class {\bf NP $\cap$ co--NP} cannot be
solved on a quantum Turing machine in time $o(2^{n/3})$.''
\end{quote} 
Is $\sqrt{N}$ speedup like in Grover algorithm the best possible result? 
It is not quite so --- in the same paper \cite{BBBV97}, we can read
\begin{quote}
\small
``What is the relevance of these oracle results?
We should emphasize that they do not rule out the possibility
that {\bf NP $\subseteq$ BQP}.
What these results do establish is that
there is no black-box approach to solving {\bf NP}-complete problems by using some
uniquely quantum-mechanical features of QTMs.''
\end{quote}
The failure with NP-complete problems is some attribute of unitary
quantum evolution, for example even small nonlinearity may help resolve
such problems \cite{NLl}. Really it is even possible to use some linear,
but nonunitary physical models, let us consider for example evolution described by
diagonal matrix with all units, except of one element $1+\epsilon$.

The model with path sums uses unitary quantum evolution. Why does it produce 
fast search, if to take into account pessimistic experience mentioned above? 
First observation is that every day we can see light travelling just along 
minimal paths. This argument may be disputed using already mentioned example 
with diffraction: such path optimization ``algorithm'' uses special structure of 
initial data, but not general one suggested in NP-complete tasks, for example it 
is smooth density for finding shortest path, or periodic structures for period 
finding discussed earlier.

Even with such limitations algorithm devotes attention, but there is yet another
argument. We may always choose period of oscillations small enough, to resolve
``diffraction problem'' for given task. Furthermore, even with diffraction such
a quantum algorithm anyway would ``perform summation'' along all paths and classical
computer need exponential time for such a task with exponential number of trajectories
or even infinite one, if we consider ``smooth path limit'' $\Delta t \to 0$, then sums 
produce famous functional integral.

\begin{figure}[htb]
\begin{center}
\parbox[t]{0.45\textwidth}{
\includegraphics[scale=0.75]{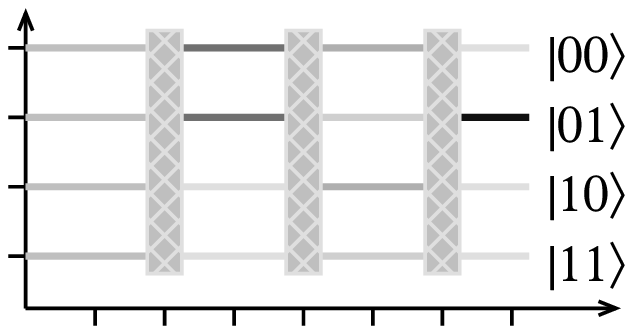}
{\caption{Quantum parallelism}
\label{fig:parall}}
}~
\parbox[t]{0.45\textwidth}{
\includegraphics[scale=0.75]{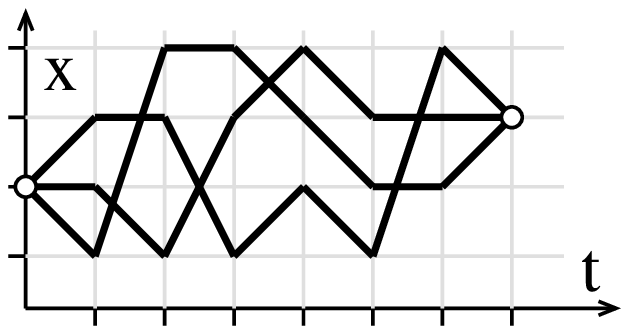}
{\caption{Feynman's paths} 
\label{fig:paths}}
}
\end{center}
\end{figure}

It is necessary to emphasize, that Feynman path integral (sum) model has certain 
difference with simpler Deutsch' quantum parallelism idea \cite{Deu85} initially
inspired by Many-Worlds Interpretation of quantum mechanics.
On Figures \ref{fig:parall} and \ref{fig:paths} depicted formal difference
between quantum parallelism and paths sum paradigm. 
This rather simple, but unexpected fact may be realized by direct analysis
of Lagrangian model described in \cite{QLagr},
but using rather rough sketch of idea, it may be 
compared with complexity of multiplication of arbitrary large (infinite in 
continuous limit) number of giant matrices in Heisenberg picture of path model 
and parallel processing with huge number of amplitudes in Schr\"odinger picture
of quantum parallelism model. In path model data is ``encoded'' in operator 
(matrix) and in Deutsch' approach --- in state (vector).

In quantum mechanics of system with infinite-dimensional Hilbert spaces
the correspondence between operators and classical function is just
subject of {\em quantization procedure}. For example classical Hamiltonian
of harmonic oscillator
\begin{equation}
 H(p,q) = p^2 + q^2
\label{Harm}
\end{equation}
corresponds to operator
\begin{equation}
 \op H = \op q^2 + \op p^2 = x^2 - \frac{\partial^2}{\partial x^2}, \quad
 \op H \colon \psi(x) \mapsto x^2 \psi(x) - \psi''(x).
\label{opHarm}
\end{equation}
How to suggest similar analogue for discrete system? Operators here
is simply matrices, but that particular matrix can be associated
with some classical function of $p$ and $q$? It is possible and
discussed in \cite{QLagr}. 

A classical function $H(p,q)$ after such quantization procedure corresponds to
some Hamiltonian, {\em e.g.} for finite-dimensional Hilbert space with dimension
$D$, it is $D \times D$ Herimtian matrix $\op H$
\begin{equation}
 \boldsymbol{\mathcal Q} \mbox{(quantization)} : 
 \quad H(p,q) \longleftrightarrow \op H.
\end{equation}
Evolution of quantum system is described by unitary matrix
\begin{equation}
 \op U(T) = e^{-i \op H T}.
\end{equation}

On the other hand, amplitude of transition between states
$\ket{k}$ and $\ket{j}$
\begin{equation}
 \bra{k} \op U(T) \ket{j} = \op U_{jk}
\end{equation}
may be expressed using sum on all paths with exponent of some expression resembling 
{\em action} in Hamiltonian mechanics with {\em classical function} $H(p,q)$  
\cite{QLagr}. It is the more precise, the more fine division of time interval 
\begin{equation}
 \Delta t = T/N, \quad U = \bigl(e^{-i \op H \Delta t}\bigr)^N.
\end{equation}

\medskip

It is clear, a model with quantum oracles and Turing machines may be not quite relevant 
for present discussion, and it is not clear, if the path integral may really resolve 
some NP-complete, exponential or even undecidable problems\footnote{Problems that 
may not be resolved by classical Turing machine in finite time, like {\em Turing 
Machine Halting Problem, Words Problem, Tiling Problem} \cite{PenEmp}, {\em etc.}} in 
theory of algorithms, or would be useful only for simulation of quantum systems.


\begin{thebibliography}{9}
\bibitem{QLagr} A. Yu. Vlasov, ``Towards Lagrangian approach to quantum
 computations,'' {\tt quant-ph/0308017}.
\bibitem{Cer} V. \v{C}ern\'y, 
``Quantum computers and intractable (NP-complete) computing problems,'' 
{\em Phys.\ Rev.\ A \bf 48}, 116--119 (1993).
\bibitem{LaurAI} J.-L. Lauri\`ere, {\em Intelligence artificielle},
(Eyrolles, Paris 1987).
\bibitem{QED} R. P. Feynman, {\em QED --- the strange theory of light and matter},
 (Princeton University Press, 1985).
\bibitem{Bennett} C. H. Bennett, ``Quantum information and computation,''
{\em Phys. Today} {\bf 48}, 24--30 (1995).
\bibitem{BBBV97} C. H. Bennett, E. Bernstein, G. Brassard, and U. Vazirani,
``Strengths and weaknesses of quantum computing,''
{\em SIAM J. Comput.} {\bf 26}, 1510--1523 (1997); {\tt quant-ph/9701001.}
\bibitem{NLl} D. S. Abrams and S. Lloyd,
``Nonlinear quantum mechanics implies poly\-nomial-time solution for
$NP$-complete and $\#P$ problems,''
{\em Phys. Rev. Lett.} {\bf 81}, 3992--3995 (1998);
{\tt quant-ph/9801041}.
\bibitem{Deu85} D. Deutsch, ``Quantum theory, the Church-Turing principle and the 
 universal quantum computer,'' {\em Proc.\ R.\ Soc.\ London Ser.\ \bf A 400}, 
 97--117 (1985).
\bibitem{PenEmp} R. Penrose, {\em The emperor's new mind}, (Oxford
  University Press, 1989)
\end{thebibliography}
\end{document}